\documentclass{article}
\usepackage{spconf,amsmath,graphicx}

\title{NeuralKalman: A Learnable Kalman Filter for Acoustic Echo Cancellation}
\name{Yixuan Zhang$^{1*}$, Meng Yu$^2$, Hao Zhang$^2$, Dong Yu$^2$, DeLiang Wang$^1$ \thanks{$^{*}$This work was done during an internship at Tencent AI Lab.}}
\address{
  $^1$The Ohio State University, Columbus, OH, USA \\
  $^2$Tencent AI Lab, Bellevue, WA, USA}
\copyrightnotice{979-8-3503-0689-7/23/\$31.00~\copyright2023 IEEE}
\begin{document}
\maketitle
 
\begin{abstract}
The robustness of the Kalman filter to double talk and its rapid convergence make it a popular approach for addressing acoustic echo cancellation (AEC) challenges. However, the inability to model nonlinearity and the need to tune control parameters cast limitations on such adaptive filtering algorithms. In this paper, we integrate the frequency domain Kalman filter (FDKF) and deep neural networks (DNNs) into a hybrid method, called NeuralKalman, to leverage the advantages of deep learning and adaptive filtering algorithms. Specifically, we employ a DNN to estimate nonlinearly distorted far-end signals, a transition factor, and the nonlinear transition function in the state equation of the FDKF algorithm. Experimental results show that the proposed NeuralKalman improves the performance of FDKF significantly and outperforms strong baseline methods.
\end{abstract}

\section{Introduction}
\label{sec:intro}

Acoustic echo cancellation (AEC), as an active and challenging research problem in the domain of speech processing, has been studied for decades and is widely used in mobile communication and teleconferencing systems. 
The goal of AEC is to eliminate the far-end signal from the near-end microphone signal so as to remove the echo of the far-end signal (back to the far end).  
In conventional digital signal processing (DSP)-based adaptive filtering algorithms \cite{duttweiler2000proportionate, gay2000fast, mader2000step, malik2012state, mohammed2007efficient} including normalized least mean square (NLMS) and affine projection, RLS, echo removal is achieved by constantly estimating the linear transfer function between the loudspeaker playing the far-end signal and the near-end microphone, known as the echo path.
However, in such AEC algorithms, control parameters need to be tuned to ensure fast convergence, and nonlinearity modeling (i.e. nonlinearity introduced by a loudspeaker) is missing. 

With recent advances in deep neural networks, deep learning-based methods \cite{zhang2018deep, zhang2022neural, yu2022neuralecho} have been utilized for AEC, and their ability to model nonlinear relations leads to promising results, even in challenging noisy or double-talk scenarios. Such methods usually treat AEC as a source separation problem and directly estimate the near-end signal based on the microphone and far-end reference signal. 
In recent AEC challenges \cite{cutler2021interspeech}, two-stage hybrid systems \cite{valin2021low, wang2021weighted, peng2021acoustic, haubner2021synergistic} that use DNN as a nonlinear post-processor of a DSP-based adaptive filtering algorithm have shown promising results. In such hybrid systems, DNNs perform nonlinear residual echo suppression, which compensates for the drawbacks of adaptive filtering algorithms. To further leverage the advantages of DNN and adaptive filtering algorithms, method such as Deep Adaptive AEC \cite{zhang2022deep} trains a hybrid model where a linear algorithm is embedded as differentiable layers, which has been proven to be highly effective in modeling a time-varying echo path.


As an adaptive filtering algorithm for AEC, the frequency domain Kalman filter (FDKF) \cite{yang2017frequency,enzner2006frequency} shows robustness in double-talk scenarios and better convergence rates. Hybrid methods based on the Kalman filter algorithm \cite{millidge2021neural, revach2022kalmannet, coskun2017long} have been used in research fields such as pose estimation, and speech filtering, but have not been well explored in the domain of AEC. The most related study is the Neural Kalman Filtering proposed in \cite{yang2022low}, where a DNN is trained to estimate a Kalman gain. 
Directly estimating the Kalman gain, however, omits crucial steps in the Kalman filter and leads to a hybrid model that resembles estimating a step size in NLMS algorithms, such as the Deep Adaptive AEC approach proposed in \cite{zhang2022deep}. Therefore, determining the optimal approach to leverage the Kalman filter and utilize DNNs to enhance the hybrid model remains an uncertain problem that is worth further investigation.

Our objective in this study is to develop a hybrid model that maximizes the benefits of both the frequency-domain Kalman filtering algorithm and DNNs. Our findings suggest that solely estimating components in the Kalman filter with DNNs does not necessarily result in improved performance. However, using DNNs to estimate missing or approximated components in the Kalman filter can lead to significant improvements. Specifically, we utilize DNN to estimate the nonlinearly distorted far-end signal, the transition factor and a nonlinear transition function in the state equation of the frequency-domain Kalman filter. Experimental results show that modeling the nonlinear distortion in far-end signals yields substantial improvements to the NeuralKalman. The transition factor shows adaptations to abrupt echo path changes and introducing a nonlinear transition function in the state equation accelerates training. Compared to modeling the covariance of the state noise and observation noise, we observe that injecting a nonlinear transition function in the state equation achieves similar improvement with less computation. The results show that the proposed hybrid NeuralKalman model suppresses echo well and outperforms the recent NLMS-based Deep Adaptive AEC \cite{zhang2022deep}. 

\section{Proposed Method: NeuralKalman}
\label{sec:method}
In a typical acoustic echo scenario, the far-end signal $x(t)$ is transmitted to the near end via a loudspeaker and received by a microphone as acoustic echo $d(t)$:
\begin{equation}
\begin{aligned}
    d(t)= h(t) * NL(x(t))
\end{aligned}
\end{equation}
where $h(t)$ represents the echo path, $NL(\cdot)$ represents the nonlinear distortion from the loudspeaker, $*$ denotes convolution. The microphone signal $y(t)$ is composed of echo $d(t)$, near-end speech $s(t)$ and noise $n(t)$:
\begin{equation}
\begin{aligned}
    y(t) = s(t) + n(t) + d(t)
\end{aligned}
\end{equation}
and it is usually processed, with the far-end signal $x(t)$ as a reference, for echo removal before being sent to the far end.

\subsection{Frequency-domain Kalman Filter}
\label{sec:fdkalman}
\begin{figure}
    \centering
    \includegraphics[width=8cm]{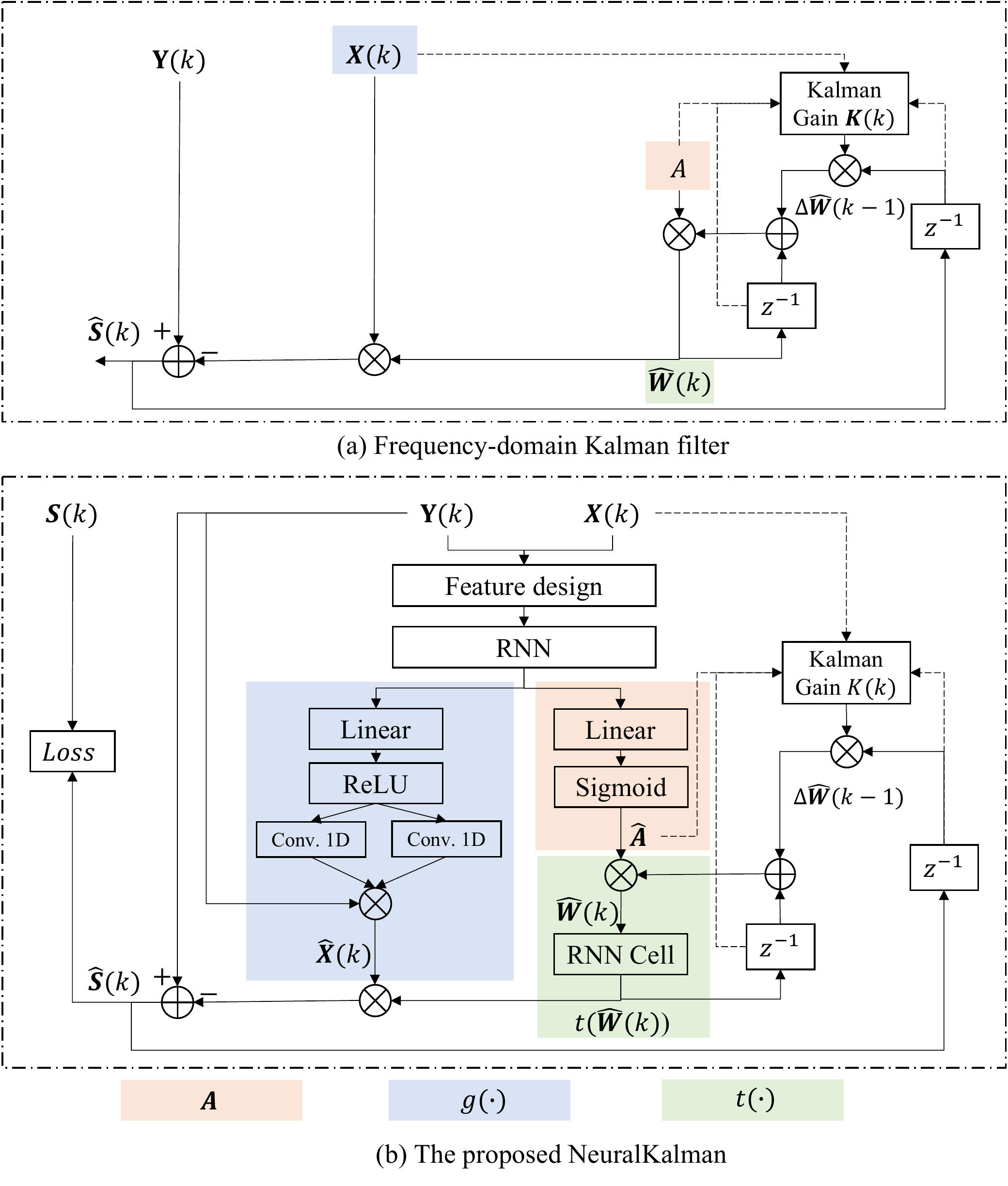}
    \caption{Diagrams of (a) Frequency-domain Kalman filter and (b) proposed NeuralKalman, where $z^{-1}$ denotes the unit delay.}
    \label{fig:fdkalman}
\end{figure}

Frequency-domain Kalman filter for AEC \cite{yang2017frequency,enzner2006frequency} estimates echo signal by modeling the echo path with an adaptive filter $\mathbf{W}(k)$ where $k$ denotes the frame index, as shown in Fig. \ref{fig:fdkalman}(a). In this study, we focus on AEC in clean condition and aim at estimating the near-end speech $s(t)$. FDKF can be interpreted as a two-step procedure and the updating of filter weights is achieved through the iterative feedback from the two steps. Following the notation in \cite{yang2017frequency}, in the prediction step, the frequency-domain near-end signal vector $\mathbf{S}(k)$ is estimated by the measurement equation, 
\begin{equation}
    \hat{\mathbf{S}}(k) = \mathbf{Y}(k) - \mathbf{G}^{01}\mathbf{X}(k)\hat{\mathbf{W}}(k),
\end{equation}
where $\mathbf{X}(k)$ is the frequency-domain far-end signal matrix, $\mathbf{Y}(k)$ corresponds to the frequency-domain microphone signal vector. $\mathbf{G}^{01}$ is the overlap-save projection matrix. $\hat{\mathbf{W}}(k)$ denotes the estimated echo path in the frequency domain. In the update step, the state equation for updating echo path $\hat{\mathbf{W}}(k)$ is defined as,
\begin{equation}
    \label{eq:state}
    \hat{\mathbf{W}}(k+1) = A[\hat{\mathbf{W}}(k)+\mathbf{G}^{01}\mathbf{K}(k)\hat{\mathbf{S}}(k)],
\end{equation}
where $A$ is the transition factor. $\mathbf{K}(k)$ denotes the Kalman gain. As shown in Fig. \ref{fig:fdkalman}(a), $\mathbf{K}(k)$ is related to far-end signal $\mathbf{X}(k)$, echo path $\hat{\mathbf{W}}(k-1)$ and estimated near-end signal $\hat{\mathbf{S}}(k-1)$. The dash line indicates the relations not expressed directly in the equations, e.g., $\hat{\mathbf{S}}(k)$ in $\mathbf{\Psi}_{vv}(k)$ \cite{yang2017frequency}. The calculation of $\mathbf{K}(k)$ is defined as,
\begin{equation}
    \mathbf{K}(k) = \mathbf{P}(k)\mathbf{X}^H(k)[\mathbf{X}(k)\mathbf{P}(k)\mathbf{X}^H(k)+2\mathbf{\Psi}_{vv}(k)]^{-1},
\end{equation}
\begin{equation}
    \mathbf{P}(k+1) = A^2[\mathbf{I}-\frac{1}{2}\mathbf{K}(k)\mathbf{X}(k)]\mathbf{P}(k) + \mathbf{\Psi}_{\Delta\Delta}(k),
\end{equation}
 where $\mathbf{P}(k)$ is the state estimation error covariance. $\mathbf{\Psi}_{vv}(k)$ and $\mathbf{\Psi}_{\Delta\Delta}(k)$ are observation noise covariance and process noise covariance respectively and are approximated by the covariance of the estimated near-end signal $\mathbf{\Psi}_{\hat{s}\hat{s}}(k)$ and the echo-path $\mathbf{\Psi}_{\hat{W}\hat{W}}(k)$, respectively. More details can be found in \cite{yang2017frequency}.


\subsection{NeuralKalman Framework}
\label{sec:kalmannet}

While being robust to double-talk and achieving a better convergence rate, the FDKF algorithm still faces several challenges. First, the echo is modeled as a linear transform of far-end signal $\mathbf{X}(k)$ while neglecting the nonlinear distortion caused by amplifiers. Second, in FDKF algorithm, $\mathbf{\Psi}_{vv}(k)$ and $\mathbf{\Psi}_{\Delta\Delta}(k)$ are approximated. An inaccurate estimate of covariance will degrade the performance of FDKF algorithm \cite{yang2017frequency}. Third, in the FDKF algorithm, the transition factor $A$ in the state equation (Eq. \ref{eq:state}) is typically assigned a fixed value that is manually adjusted based on the non-stationarity of the echo path. However, a fixed $A$ is less likely to adapt well to the changing environment.

To address these problems, we propose NeuralKalman. Unlike \cite{yang2022low} which directly estimates Kalman gain from concatenated input features comprising estimated far-end, near-end signals, and innovation of $\hat{W}$, the proposed framework leverages DNNs to estimate the transition factor $A$, the far-end nonlinear distortion $g(\cdot)$, and a nonlinear transition function $t(\cdot)$, as shown in Figure \ref{fig:fdkalman}(b). 
Since the echo path and nonlinear distortion information can be retrieved from the microphone and far-end signals, the input feature computed from the complex STFT of the microphone signal and far-end signal is shared for estimating transition factor $A$ and nonlinear distortion $g(\cdot)$. Similar to NeuralEcho \cite{yu2022neuralecho}, the employed input feature is a concatenation of temporal correlation, frequency correlation, channel covariance, and normalized log power spectrum of microphone and far-end signal. 
As shown in Fig. \ref{fig:fdkalman}(b), an recurrent neural network (RNN) takes the computed input feature and is followed by two branches for estimating nonlinearly distorted far-end signal $\hat{\mathbf{X}}(k)$ and transition factor $\hat{A}(k)$ respectively. The shared RNN is a 4-layer long short-term memory (LSTM) network where each layer has 257 hidden units.
 
\subsubsection{Nonlinear Distortion $g(\cdot)$}
To address the nonlinear distortion introduced by the loudspeaker, we estimate the far-end nonlinear distortion with DNNs and use the nonlinear far-end signal as a reference for updating the Kalman filter. As illustrated in Figure \ref{fig:fdkalman}(b), the sub-network responsible for estimating $g(\cdot)$ consists of a linear layer with Rectified Linear Unit (ReLU) activation, followed by two one-dimensional convolution layers (Conv. 1D). The nonlinearly distorted far-end signal $\hat{\mathbf{X}}(k)$ is obtained by applying the complex-valued ratio filters $cRF$ \cite{mack2019deep, yu2022neuralecho} to the microphone signal $\mathbf{Y}(k)$. 

\subsubsection{Nonlinear Transition Function $t(\cdot)$}
In \cite{yang2022low}, the paper illustrates the effectiveness of implicitly modeling the covariance of state noise and observation noise during the Kalman gain estimation process. Our observation, as discussed in Sec. \ref{sec:expresults}, also demonstrates that incorporating covariance modeling enhances performance. However, we find that introducing a nonlinear transition function in the state equation leads to a comparable performance improvement with less computation. More specifically, we replace the linear transition function in Eq. \ref{eq:state} with a nonlinear one:
\begin{equation}
\label{eq:nlstate}
\hat{\mathbf{W}}(k+1) = t(A[\hat{\mathbf{W}}(k)+\mathbf{G^{01}}\mathbf{K}(k)\hat{\mathbf{S}}(k)]),
\end{equation}
where the nonlinear transition function $t(\cdot)$ is estimated from an LSTM cell which has 256 hidden units. The input to the LSTM cell consists of the estimated $\hat{\mathbf{W}}(k+1)$ from Eq. \ref{eq:state} and previous state $h_{k-1}$. Then two linear layers take $h_{k}$ as input and output the real and imaginary parts of the processed $t(\hat{\mathbf{W}}(k+1))$.
\begin{equation}
\begin{aligned}
       h_{k} = \mathbf{RNN}(\hat{\mathbf{W}}(k+1), h_{k-1}), \\
    t(\hat{\mathbf{W}}(k+1)) = \mathbf{FNN}(h_{k}), 
\end{aligned}
\end{equation}
where $\mathbf{RNN}$ and $\mathbf{FNN}$ denote the LSTM cell and linear layers respectively.

\subsubsection{Transition Factor $A$}
\label{ssec:transitionfactor}

Transition factor $A$ in the range of [0,1] depicts the variation of the Kalman filter and it is often manually tuned to a value that is close to 1. To incorporate the influence of possible changes in the echo path on the transition factor, instead of using a fixed value, we employ DNN to estimate a time-varying transition factor for Eq. \ref{eq:state}.
 The branch for estimating frame-based transition factor $A(k)$ is composed of a linear layer followed by a sigmoidal activation function. 


\subsubsection{Loss Function}

The loss function is defined to jointly optimize SI-SDR \cite{le2019sdr} in the time domain and mean absolute error (MAE) of magnitude spectrogram between the target and estimated near-end signal.
\begin{equation}
    L = -\mathbf{SI\text{-}SDR}(s, \hat{s}) + \alpha \mathbf{MAE}(|\mathbf{S}|, |\hat{\mathbf{S}}|),
\end{equation}
where $\hat{s}$ and $\hat{\mathbf{S}}$ are the estimated time-domain and frequency-domain near-end signal, respectively. And $\alpha$ is set to $10,000$ in our implementation to balance the value range of the two losses.

\section{Experimental setup}
\label{sec:expsetup}

\subsection{Dataset}

Following \cite{yu2022neuralecho}, we simulate the single-channel AEC dataset using AISHELL-2 \cite{du2018aishell} and AEC-Challenge \cite{cutler2021interspeech} datasets. We use clean and nonlinearly distorted far-end signals from AEC-Challenge's synthetic echo set \cite{cutler2021interspeech}. Nonlinear distortions such as maximum amplitude clipping with a Sigmoidal function \cite{zhang2018deep}, learned distortion functions, etc. are included in the far-end signals. To simulate acoustic echo, 10k room impulse responses (RIRs) sets with random room characteristics are generated using the image-source method \cite{allen1979image} with reverberation time (RT60) ranging from 0 to 0.6 seconds. Each of the 10k RIRs sets comprises the RIRs from locations of the loudspeaker, near-end speaker. During data generation, the signal-to-echo-ratio (SER) ranges from -10 dB to 10 dB and RIRs set is randomly picked. The training set has 90k utterances, and 10k utterances are randomly selected in each epoch for training. Each network is trained for 90 epochs. We generated 200 utterances for validation and 300 utterances for testing. The test set is generated from utterances and RIRs that are not seen in training process. All input audios are sampled at 16 kHz. STFT is computed with a 32 ms frame length and 50\% frame shift.

\subsection{Evaluation Metrics}

We evaluate the echo cancellation performance of the proposed NeuralKalman using perceptual evaluation of speech quality (PESQ) \cite{rix2001perceptual} and word error rate (WER). 
To evaluate WER, we use a commercial general-purpose speech recognition API \cite{tencentasr} to test the automatic speech recognition (ASR) performance.


\section{Experimental Results}
\label{sec:expresults}

\subsection{NeuralKalman Evaluation}

\begin{table}[]
\caption{Performance of NeuralKalman with various settings in the presence of double-talk.}
\label{tab:kalmannet}
\centering
\begin{tabular}{l|cc} \hline
                & PESQ & WER     \\  \hline
Unprocessed     & 1.87 & 79.85\%    \\ \hline
Kalman Filter \cite{enzner2006frequency}     & 2.32  & 32.89\% \\
NeuralKalman-$g(\cdot)/t(\cdot)/A$ & \textbf{2.67} & \textbf{20.41\%}  \\
NeuralKalman-$g(\cdot)/A$   & 2.57 & 22.37\%  \\
NeuralKalman-$g(\cdot)/t(\cdot)$ & 2.59 & 21.98\% \\
NeuralKalman-$g(\cdot)/\mathbf{\Psi}/A$ & 2.65 & 21.79\% \\
NeuralKalman-$g(\cdot)/\mathbf{\Psi}/t(\cdot)/A$ & \textbf{2.67} & 20.72\% \\
\hline
\end{tabular}
\end{table}

We conduct an ablation study to primarily investigate the impact of modeling the nonlinear transition function $t(\cdot)$ and the transition factor $A$. The influence of modeling of nonlinear distortion $g(\cdot)$ has been well examined in prior research \cite{zhang2022deep}. NeuralKalman models with different DNN components are built for comparison and the results are shown in Table \ref{tab:kalmannet}. 
The NeuralKalman model discussed in Sec. \ref{sec:kalmannet} and shown in Fig. \ref{fig:fdkalman}(b) corresponds to NeuralKalman-$g(\cdot)/t(\cdot)/A$ which estimates a nonlinear distortion function $g(\cdot)$, a nonlinear transition function $t(\cdot)$ and the transition factor $A$. NeuralKalman-$g(\cdot)/A$ uses DNN to jointly estimate a nonlinear distortion function $g(\cdot)$ and the transition factor $A$. NeuralKalman-$g(\cdot)/t(\cdot)$ estimates a nonlinear distortion function $g(\cdot)$ and a nonlinear transition function $t(\cdot)$.

Since an accurate estimation of covariance matrices in FDKF would contribute to better convergence rate and AEC performance \cite{yang2017frequency}, we also train a NeuralKalman-$g(\cdot)/\mathbf{\Psi}/A$ model which involves training two LSTM cells with 256 hidden units to estimate the covariance matrices $\mathbf{\Psi}_{vv}(k)$ and $\mathbf{\Psi}_{\Delta\Delta}(k)$. The inputs to the RNNs for estimating $\mathbf{\Psi}_{vv}(k)$ and $\mathbf{\Psi}_{\Delta\Delta}(k)$ are the estimated near-end speech $\hat{\mathbf{S}}(k)$ and updated $\hat{\mathbf{W}}(k)$, respectively. NeuralKalman-$g(\cdot)/\mathbf{\Psi}/t(\cdot)/A$ which performs both covariance matrices and nonlinear transition function estimation is also trained for comparison. 

\subsubsection{Nonlinear Distortion $g(\cdot)$}
While the influence of modeling $g(\cdot)$ has been discovered in \cite{zhang2022deep}, we also observe that the key improvement comes from modeling the far-end nonlinear distortion. We find that by solely estimating $A$, we achieve a PESQ score of 2.32 and a WER of 31.67\% which is slightly better than Kalman filter. Compared to only estimating $A$, we find that further introducing far-end nonlinear distortion $g(\cdot)$ increases PESQ by 0.25, and reduces WER relatively by 29.4\%.

\subsubsection{Nonlinear Transition Function $t(\cdot)$}

\begin{figure}[]
\centering
\includegraphics[width=5.0cm]{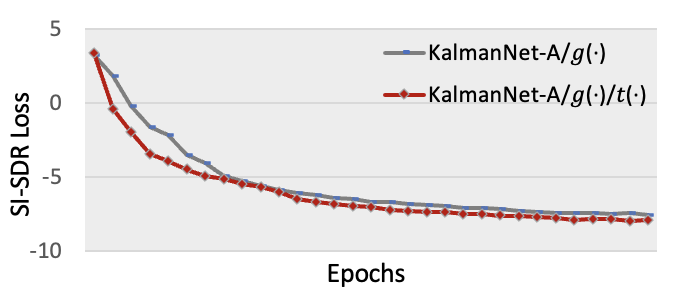}
\caption{SI-SDR loss curve of NeuralKalman-$A/g(\cdot)$ and NeuralKalman-$A/g(\cdot)/t(\cdot)$.}
\label{fig:trainingloss}
\end{figure}

With the learned nonlinear transition function $t(\cdot)$, the performance of NeuralKalman is further improved. As shown in Table \ref{tab:kalmannet}, by comparing the results of NeuralKalman-$g(\cdot)/A$ and NeuralKalman-$g(\cdot)/t(\cdot)/A$, we find that PESQ is improved by 0.1, and WER is relatively reduced by 8.8\%. NeuralKalman-$g(\cdot)/\mathbf{\Psi}/A$ which substitute approximated $\mathbf{\Psi}_{vv}(k)$ and $\mathbf{\Psi}_{\Delta\Delta}(k)$ in FDKF with DNN learned covariance, has also shown improved performance in terms of all metrics. 
Compared to NeuralKalman-$g(\cdot)/\mathbf{\Psi}/A$, we observe that NeuralKalman-$g(\cdot)/t(\cdot)/A$ can achieve slightly better performance with less computational cost. Also, the result of NeuralKalman-$g(\cdot)/\mathbf{\Psi}/t(\cdot)/A$ shows that estimating both covariance and nonlinear transition function does not bring further improvement. In addition, we observe from Fig. \ref{fig:trainingloss}  that estimating the nonlinear transition function $t(\cdot)$ brings faster training convergence speed. Therefore, we decide to only estimate $t(\cdot)$ instead of $\mathbf{\Psi}$. 

\subsubsection{Transition Factor $A$}
\begin{figure}[]
\centering
\includegraphics[width=8cm]{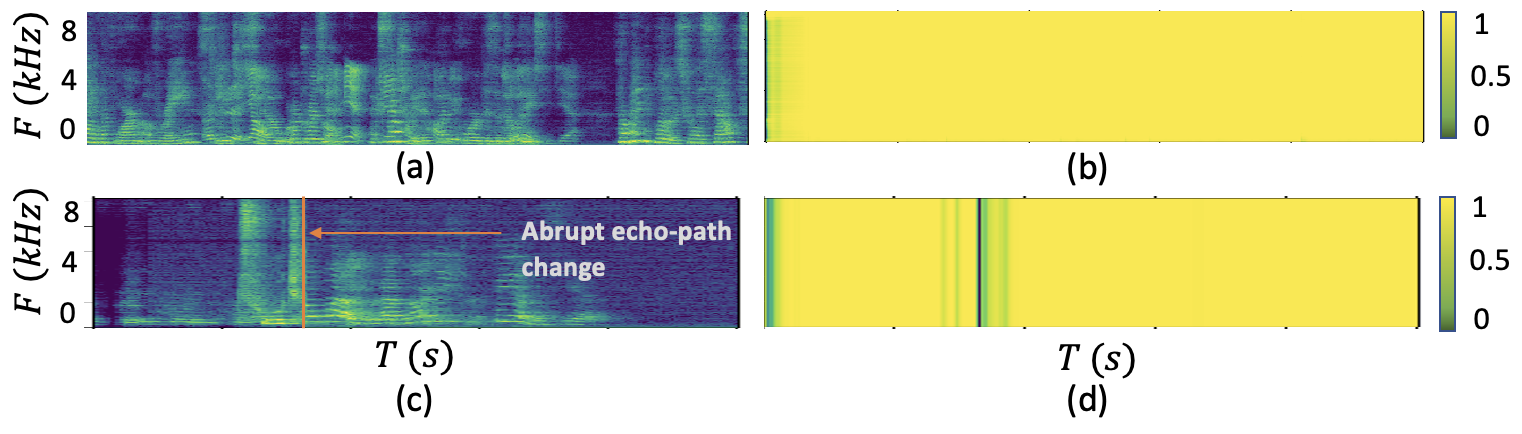}
\caption{Estimated transition factor $A$ for signals with no echo path change (a, b) and with abrupt echo path change (c, d): (a)(c) Magnitude STFT of a microphone signal, (b)(d) Estimation of $A$ from NeuralKalman-$A$ model. }
\label{fig:A}

\end{figure}

By incorporating the modeling of the transition factor $A$, the Kalman filter gains enhanced flexibility in controlling the update of the echo path We validate the necessity of estimating transition factor $A$ by eliminating the estimation of $A$ from the best-performing NeuralKalman-$g(\cdot)/t(\cdot)/A$ model, where we find that PESQ reduces by 0.1 and WER increases to $21.98\%$.
To examine the response of $A$ in the scenario of abrupt echo path change, we have also evaluated a model that solely estimates $A$ (named as NeuralKalman-$A$) on a sample with abrupt echo path change. The RIR is aburptly switched from one to another at the indicated position in Fig. \ref{fig:A}. We plot the value of learned $A$ from the NeuralKalman-$A$ model on audios with and without abrupt echo path change. 
It is observed that for the audio signal without echo path change, the learned $A$ is close to 1 and relatively stable throughout time, which explains why the performance of NeuralKalman-$A$ is similar to that of FDKF. For the signal with abrupt echo path change, we observe that the value of learned $A$ decreases to nearly 0 when the echo path abruptly changes and gradually increases afterward, which is reasonable to obtain a stable convergence. We believe that it is proper to use a time-varying $A$ in scenarios with echo path changes to make the updating of the algorithm stable and diminish the chances of divergence. 

\subsubsection{Other observations}

Other experiments such as using additional LSTM cells to estimate the complex matrices of $\mathbf{K}(k)$ and $\mathbf{P}(k)$ on top of NeuralKalman-$g(\cdot)/t(\cdot)/A$ are performed and achieve similar performance. It is observed that solely estimating Kalman filter components using DNNs did not consistently improve performance. However, estimating missing or approximated components yields significant improvements.

\subsection{Comparison with Baselines}
\begin{table}[]
\caption{Performance comparison with baseline methods.}
\label{tab:comparison}
\centering
\begin{tabular}{l|cc}
\hline
                & PESQ & WER    \\  \hline
Unprocessed     & 1.87 & 79.85\%      \\ \hline
Kalman Filter \cite{enzner2006frequency}   & 2.32 & 32.89\%   \\
NLMSNet \cite{zhang2022deep}\footnotemark   & 2.52 & 23.33\% \\
DNN-AEC   & 2.62 & 23.14\%  \\
NeuralKalman-$g(\cdot)/t(\cdot)/A$ & \textbf{2.67} & \textbf{20.41\%}  \\
\hline
\end{tabular}
\end{table}

\begin{figure}[]
\centering
\includegraphics[width=8.5cm]{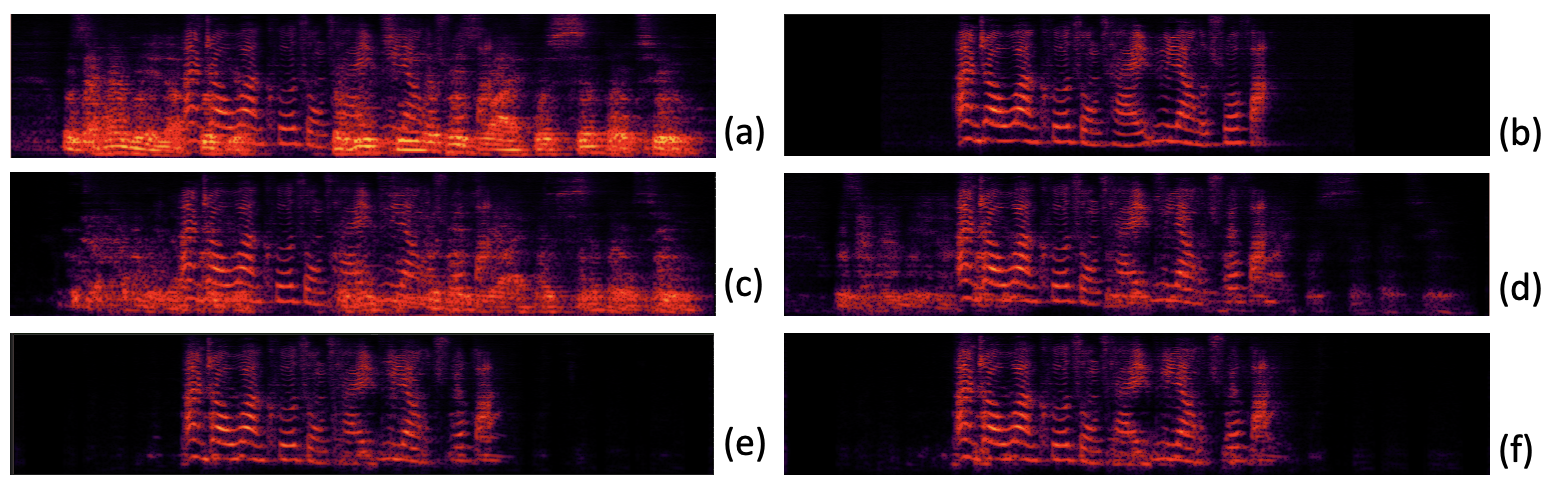}
\caption{Spectrograms of (a) microphone signal, (b) target near-end signal, and outputs of (c) Kalman filter, (d) NLMSNet, (e) DNN-AEC, and (f) our proposed NeuralKalman-$g(\cdot)/t(\cdot)/A$.}
\label{fig:spectrogram}
\end{figure}

We compare the proposed NeuralKalman-$g(\cdot)/t(\cdot)/A$ model with strong baseline methods including frequency-domain Kalman filter \cite{enzner2006frequency}, NLMSNet which is based on deep adaptive AEC \cite{zhang2022deep} \footnotetext{NLMSNet is a modified and retrained version of \cite{zhang2022deep} for fair comparison.}, and a fully DNN-based model DNN-AEC. Following \cite{zhang2022deep}, NLMSNet is a hybrid model based on NLMS algorithm and takes microphone and far-end signal as inputs and estimate the step size parameter and the non-linear far-end signal. DNN-AEC is a method based entirely on DNNs that uses the microphone and far-end signal as inputs to directly predict the speech at the near-end. In our experiments, NLMSNet and DNN-AEC adopt the same RNN network, input feature and loss function as the proposed NeuralKalman which are described in Sec. \ref{sec:kalmannet}. 
From Table \ref{tab:comparison}, we observe that all hybrid methods outperform the frequency-domain Kalman filter algorithm. Among the hybrid methods, NeuralKalman has the best performance. We observe that NLMSNet does not show superiority over DNN-AEC when trained on data with stationary echo path. Compared to NLMSNet, NeuralKalman improves the PESQ by 0.15 and relatively improves WER by 12.5\%. NeuralKalman outperforms DNN-AEC in terms of PESQ and WER, with PESQ showing a 0.05 improvement and WER showing a relative improvement of 11.8\%. 
Fig. \ref{fig:spectrogram} shows the magnitude STFT of the near-end signal estimated by different methods. 
The frequency-domain Kalman filter's ability to suppress echoes appears to be limited. Hybrid method such as NLMSNet shows promising echo suppression results in double talk regions; however, echoes remain present in single talk regions. Among the baseline methods, DNN-AEC and NeuralKalman-$g(\cdot)/t(\cdot)/A$ demonstrate superior effectiveness in removing echoes. 


\section{Conclusion}
\label{sec:copyright}

In this paper, we have proposed a learnable Kalman filter for acoustic echo cancellation. The proposed model leverages the advantages of DNN to improve the Kalman filter by estimating the missing or approximated components, including the transition factor, nonlinear distortion of the far-end signal, and nonlinear transition function for the estimated echo path. Systematic evaluations show that the proposed method outperforms recent baseline methods. For future work, with more access to data, we will explore training NeuralKalman on real-recorded signals with echo path changes and explore utilizing it in real-world devices.

\clearpage

\bibliographystyle{IEEEbib}
\bibliography{mybib}

\begin{thebibliography}{10}

\bibitem{duttweiler2000proportionate}
Donald~L Duttweiler,
\newblock ``Proportionate normalized least-mean-squares adaptation in echo
  cancelers,''
\newblock {\em IEEE Transactions on Speech and Audio Processing}, vol. 8, pp.
  508--518, 2000.

\bibitem{gay2000fast}
Steven~L Gay,
\newblock ``The fast affine projection algorithm,''
\newblock in {\em Acoustic Signal processing for Telecommunication}, pp.
  23--45. 2000.

\bibitem{mader2000step}
Andreas Mader, Henning Puder, and Gerhard~Uwe Schmidt,
\newblock ``Step-size control for acoustic echo cancellation filters--an
  overview,''
\newblock {\em Signal Processing}, vol. 80, pp. 1697--1719, 2000.

\bibitem{malik2012state}
Sarmad Malik and Gerald Enzner,
\newblock ``State-space frequency-domain adaptive filtering for nonlinear
  acoustic echo cancellation,''
\newblock {\em IEEE Transactions on Audio, Speech, and Language Processing},
  vol. 20, pp. 2065--2079, 2012.

\bibitem{mohammed2007efficient}
Jafar~Ramadhan Mohammed and Gurnam Singh,
\newblock ``An efficient \uppercase{RLS} algorithm for output-error adaptive
  \uppercase{IIR} filtering and its application to acoustic echo
  cancellation,''
\newblock in {\em IEEE Symposium on Computational Intelligence in Image and
  Signal Processing}, 2007, pp. 139--145.

\bibitem{zhang2018deep}
Hao Zhang and DeLiang Wang,
\newblock ``Deep learning for acoustic echo cancellation in noisy and
  double-talk scenarios,''
\newblock in {\em Proc. Interspeech}, 2018, p. 322.

\bibitem{zhang2022neural}
Hao Zhang and DeLiang Wang,
\newblock ``Neural cascade architecture for multi-channel acoustic echo
  suppression,''
\newblock {\em IEEE/ACM Transactions on Audio, Speech, and Language
  Processing}, vol. 30, pp. 2326--2336, 2022.

\bibitem{yu2022neuralecho}
Meng Yu, Yong Xu, Chunlei Zhang, Shi-Xiong Zhang, and Dong Yu,
\newblock ``Neural\uppercase{E}cho: A self-attentive recurrent neural network
  for unified acoustic echo suppression and speech enhancement,''
\newblock {\em arXiv preprint arXiv:2205.10401}, 2022.

\bibitem{cutler2021interspeech}
Ross Cutler, Ando Saabas, Tanel Parnamaa, Marju Purin, Hannes Gamper, Sebastian
  Braun, Karsten S{\o}rensen, and Robert Aichner,
\newblock ``{ICASSP} 2022 acoustic echo cancellation challenge,''
\newblock in {\em Proc. ICASSP}, 2022, pp. 9107--9111.

\bibitem{valin2021low}
Jean-Marc Valin, Srikanth Tenneti, Karim Helwani, Umut Isik, and Arvindh
  Krishnaswamy,
\newblock ``Low-complexity, real-time joint neural echo control and speech
  enhancement based on percepnet,''
\newblock in {\em Proc. ICASSP}, 2021, pp. 7133--7137.

\bibitem{wang2021weighted}
Ziteng Wang, Yueyue Na, Zhang Liu, Biao Tian, and Qiang Fu,
\newblock ``Weighted recursive least square filter and neural network based
  residual echo suppression for the \uppercase{AEC}-challenge,''
\newblock in {\em Proc. ICASSP}, 2021, pp. 141--145.

\bibitem{peng2021acoustic}
Renhua Peng, Linjuan Cheng, Chengshi Zheng, and Xiaodong Li,
\newblock ``Acoustic echo cancellation using deep complex neural network with
  nonlinear magnitude compression and phase information.,''
\newblock in {\em Proc. Interspeech}, 2021, pp. 4768--4772.

\bibitem{haubner2021synergistic}
Thomas Haubner, Mhd~Modar Halimeh, Andreas Brendel, and Walter Kellermann,
\newblock ``A synergistic \uppercase{K}alman and deep postfiltering approach to
  acoustic echo cancellation,''
\newblock in {\em Proc. EUSIPCO}, 2021, pp. 990--994.

\bibitem{zhang2022deep}
Hao Zhang, Srivatsan Kandadai, Harsha Rao, Minje Kim, Tarun Pruthi, and Trausti
  Kristjansson,
\newblock ``Deep adaptive \uppercase{AEC}: Hybrid of deep learning and adaptive
  acoustic echo cancellation,''
\newblock in {\em Proc. ICASSP}, 2022, pp. 756--760.

\bibitem{yang2017frequency}
Feiran Yang, Gerald Enzner, and Jun Yang,
\newblock ``Frequency-domain adaptive \uppercase{K}alman filter with fast
  recovery of abrupt echo-path changes,''
\newblock {\em IEEE Signal Processing Letters}, vol. 24, pp. 1778--1782, 2017.

\bibitem{enzner2006frequency}
Gerald Enzner and Peter Vary,
\newblock ``Frequency-domain adaptive \uppercase{K}alman filter for acoustic
  echo control in hands-free telephones,''
\newblock {\em Signal Processing}, vol. 86, pp. 1140--1156, 2006.

\bibitem{millidge2021neural}
Beren Millidge, Alexander Tschantz, Anil Seth, and Christopher Buckley,
\newblock ``Neural \uppercase{k}alman filtering,''
\newblock {\em arXiv preprint arXiv:2102.10021}, 2021.

\bibitem{revach2022kalmannet}
Guy Revach, Nir Shlezinger, Xiaoyong Ni, Adria~Lopez Escoriza, Ruud~JG
  Van~Sloun, and Yonina~C Eldar,
\newblock ``Kalmannet: Neural network aided kalman filtering for partially
  known dynamics,''
\newblock {\em IEEE Transactions on Signal Processing}, vol. 70, pp.
  1532--1547, 2022.

\bibitem{coskun2017long}
Huseyin Coskun, Felix Achilles, Robert DiPietro, Nassir Navab, and Federico
  Tombari,
\newblock ``Long short-term memory kalman filters: Recurrent neural estimators
  for pose regularization,''
\newblock in {\em Proc. of the IEEE International Conference on Computer
  Vision}, 2017, pp. 5524--5532.

\bibitem{yang2022low}
Dong Yang, Fei Jiang, Wei Wu, Xuefei Fang, and Muyong Cao,
\newblock ``Low-complexity acoustic echo cancellation with neural
  \uppercase{K}alman filtering,''
\newblock {\em arXiv preprint arXiv:2207.11388}, 2022.

\bibitem{mack2019deep}
Wolfgang Mack and Emanu{\"e}l~AP Habets,
\newblock ``Deep filtering: \uppercase{S}ignal extraction and reconstruction
  using complex time-frequency filters,''
\newblock {\em IEEE Signal Processing Letters}, vol. 27, pp. 61--65, 2019.

\bibitem{le2019sdr}
Jonathan Le~Roux, Scott Wisdom, Hakan Erdogan, and John~R Hershey,
\newblock ``\uppercase{SDR}--half-baked or well done?,''
\newblock in {\em Proc. ICASSP}, 2019, pp. 626--630.

\bibitem{du2018aishell}
Jiayu Du, Xingyu Na, Xuechen Liu, and Hui Bu,
\newblock ``\uppercase{AISHELL}-2: Transforming mandarin \uppercase{ASR}
  research into industrial scale,''
\newblock {\em arXiv:1808.10583}, 2018.

\bibitem{allen1979image}
Jont~B Allen and David~A Berkley,
\newblock ``Image method for efficiently simulating small-room acoustics,''
\newblock {\em The Journal of the Acoustical Society of America}, vol. 65, pp.
  943--950, 1979.

\bibitem{rix2001perceptual}
Antony~W Rix, John~G Beerends, Michael~P Hollier, and Andries~P Hekstra,
\newblock ``Perceptual evaluation of speech quality (\uppercase{PESQ}) -
  \uppercase{A} new method for speech quality assessment of telephone networks
  and codecs,''
\newblock in {\em Proc. ICASSP}, 2001, pp. 749--752.

\bibitem{tencentasr}
``{Tencent \uppercase{ASR}},'' in http://ai.qq.com/product/aaiasr.shtml.

\end{thebibliography}

\end{document}